\documentclass[twocolumn,showpacs,preprintnumbers,amsmath,amssymb, showkeys]{revtex4}
\usepackage{revsymb}
\usepackage{graphicx}% Include figure files
\usepackage{dcolumn}% Align table columns on decimal point
\usepackage{bm}% bold math

\begin{document}

\preprint{arXiv:yymm.nnnn}%

\title{Fitting the Galaxy Rotation Curves: Strings versus NFW profile}% Force line breaks with \\

\author{Yeuk-Kwan E. Cheung and Feng Xu}

\affiliation{%
Department of Physics, Nanjing University\\
22 Hankou Road, Nanjing 210098, P. R. China
}%

\date{June 4,  2008}%{\today}% It is always \today, today,
             
% \thanks{\texttt{[cheung.edna<at>gmail.com]}}

\begin{abstract}
Remarkable fit of galaxy rotation curves is achieved using a  simple  model  from string theory.
The rotation curves  of  the same group of  galaxies are also fit using dark matter model
with  the  generalized Navarro-Frenk-White  profile   for comparison.  
String model utilizes three free parameters vs five in the  dark matter model. 
The average $\chi^2$ of the string model fit is $1.649$ while that of the dark matter model 
is $1.513$.    The generalized NFW profile  fits marginally better at a  price of  two more free parameters.

%Valid PACS numbers may be entered using the \verb+\pacs{#1}+ command.
\end{abstract}
%Use showkeys class option if keyword  display desired
\pacs{Valid PACS appear here}% PACS, the Physics and Astronomy  Classification Scheme.
\maketitle

While the Cold Dark Matter cosmology has been accepted by many as the correct theory for structure formation 
at large scale and the solution to the missing mass problem at the galactic scale
 there still lacks hard proof for the existence of Dark Matter particles.  
Until the coming Linear Hadron Collider, and future experiments,  
tells us definitely what constitute Dark Matter
a more natural and universal explanation in lieu of dark matter cannot be excluded.  
Here we entertain the possibility that a higher-rank gauge field universally coupled to strings 
can act  like a Lorentz force in four dimensions providing an extra centripetal acceleration for matter
towards the center of a galaxy.   If not properly accounted for,  
it would appear as if there were extra {\it{invisible}} matter in a galaxy.  
A salient feature of this Lorentz-like force is that it  fits  
 galaxies 
with an extended region of linearly rising rotation velocity significantly better than the dark matter model.  
This feature also endows the model with testability: in the region where the gravitational attraction of the 
visible matter completely gives way to the linear rising Lorentz force, typically in the region $ r \sim 20R_{d}$,  
should we still observe  linear rising rotation velocity, say for the satellites of the host galaxy, 
it would be a strong support for the model.  
Otherwise if  {{\it{all}}} rotation curves  are found to  fall off beyond $  20R_{d}$  then the model is proven wrong.  
Furthermore this is the first of such attempts  to verify directly the validity of string theory as a description of  low energy physics.   Given this last  reason alone we regard it a worthwhile endeavour.   

\section{The String Model}    
\label{sec: string_model}

Consider a four-dimensional string model proposed by Nappi and Witten~\cite{Nappi:1993ie} 
 in which  the string theory is exactly integrable to all orders of the string scale.   
 Furthermore all  tree-level correlation functions which capture  finite-sized effect of the strings,  
exact  to all orders in string scale, have been computed~\cite{Cheung:2003ym}. 
This  is valid for all energy scales as long as the string coupling constant is weak.  
The three-form background  gauge field,  $H_{(3)}$,  coupled uniquely to the worldsheet of the strings,  is constant.   
Because we are no longer approximating strings as point particles, this coupling between the two-form gauge potential 
$B_{(2)}$ and the two dimensional worldsheet of the string produces a net force on the string when viewed as a single particle.
The center of mass of the closed string executes Landau orbits:
\begin{equation}  \label{eq: landau}
a = a_0 + r  e^{iHt}
\end{equation}
where $a$ is the complex coordinate of the plane in which the time-like part of the three-form field has non-zero components. 
This phenomenon is analogous to   the behaviour of an electron in  a constant magnetic field.  

For the present purposes a slightly different field configuration is required as compared to the exactly solvable Nappi-Witten model.  We retain the spatially constant time-like component of the field strength $H_{0ij}$ only, denoted by $H$.  We assume that the field is aligned with the plane of the galaxy, and envision that it may be self-consistently produced by the galaxy itself via a dynamo effect.
This possibility is favoured from the string theory point of view, because the coupling of the field to matter has universal strength, i.e. all matter is charged, rather than neutral. 
Therefore if all matter is indeed made up of fundamental strings and hence couples universally to $H_{0ij}$
each star in a  galaxy  will then execute  the circular motion in concentric landau orbits on the galactic plane. 
Effectively there is an additional Lorentz force term in the equation of motion for a test star:
\begin{equation}
 \label{eq: EOM}
m\,  \frac{v^2}{r} = q\,H\,v + \,  m\, F_{*}~
\end{equation}
where the field, $H$, is generated by the rotating stellar matter and the halo of  gases alike.

To describe the visible mass we use the parametric distribution with exponential fall off in density due to  van der Kruit and Searle:
\begin{equation}        \label{eq: vdKS}
\rho  (r, z) = \rho_0\, \displaystyle{ e^{-\, \frac{r}{R_d}}}\, sech^2(\frac{z}{Z_d})
\end{equation}
with $\rho_0$ being the central matter density, 
$R_d$  the characteristic radius of the stellar disc and $Z_d$ the characteristic thickness.  
Following a common practice we choose $Z_d$ to be $\frac{1}{6} R_d$; the dependence of the final results  
on this choice is very weak.

Gravitational attraction due to the visible matter is  given by  
\begin{eqnarray}
F_{*}({r})
 &=& G_{N}\, \rho_0\, R_{d}\,  \tilde{F} (\tilde{r} \equiv \frac{r}{R_{d}}) 
\end{eqnarray}
where 
$$\tilde{F}(\tilde{r})= \frac{\partial }{\partial  \tilde{r}} \, \int_{all\, space}
\displaystyle{\frac{e^{-r'}sech^2(6z')}{|\vec{ \tilde{r}}-\vec{r'}|}}~r'dr'dz'd\phi' $$ 
is  universal function  for all galaxies.
We are now ready to fit the galaxy rotation curves data with three free parameters, $\Omega$, $R_d$, and $\rho_0$ 
defined by the following equation:
\begin{equation}                  \label{eq: eom_string}
\frac{v^2}{r}  = 2\, \Omega\,  v + G_{N} \rho_0\,  R_{d}\,  \tilde{F}  
 \end{equation}
 with  $\Omega \equiv \frac{qH}{2m}$.   
Together with the fundamental charge-to-mass ratio  the strength of the gauge field is encoded in the 
parameter $\Omega$.
 
 A few remarks are in order.
 Independent of any galaxy rotation modelling, $ \rho$ and $R_d$ can be either fit or cross checked with  
 photometric data and hence are not really free parameters.  
 Since we are only interested in comparing  the dark matter model and our string model  in  fitting 
 the galaxy rotation curves we are treating them as free parameters in {{\it{both}}} models.  
 We will later on use the best fit values for these two parameters 
 (and three others in the dark matter model) to compute the total mass, 
 as well as the mass-to-light ratios, for these galaxies.     
 These will serve as sanity check for the best fit values of the parameters in  both models.
We are ignoring the gas contribution from our fitting because putting in more free parameters will no doubt 
improve the fit for both models.   
For the same reason we do not allow for any correction for star extinctions and supernova feedback  
as they  would not affect any conclusion we draw  concerning the {{\it relative}} quality of the fit.   
Keeping this simplistic spirit  we do not allow for any dark matter component at all in the string model  
and we also  assume that the strength of the string field is constant throughout the span of each galaxy.   
Back reaction of spacetime  to the presence of the string field is also ignored.

\section{The Dark Matter Model}    
\label{sec: dark_matter_model}

According to the Cold Dark Matter (CDM) paradigm  there is approximately ten times more dark matter than visible matter.  
The  fluctuations of the primordial density perturbations of the universe get amplified by gravitational instabilities.  
Hierarchical clustering models
  furthermore   predict that dark matter density traces the density of the universe 
at the time of collapse and thus all dark matter halos have similar density.    
Baryons  then fall into the gravitational potential created by the clusters of  dark matter particles, 
forming the visible part of the galaxies. 
In a galaxy the dark matter exists in a spherical halo engulfing all of the visible matter and 
extending  much further beyond the stellar disc.  
To describe the dark matter component we use the generalized Navarro-Frenk-White profile~\cite{NFW}:
\begin{equation}        \label{eq: NFW}
\sigma =\frac{\sigma_0 } { (\frac{r}{r_s})^\alpha (1 + \frac{r}{r_s})^{3-\alpha}}~,
\end{equation}
where $\alpha=1$ corresponds to the NFW profile.  
 $r_s$ is the characteristic radius of the dark matter halo.  
In the fitting routine we allow the $r_s$ to vary from $3R_d$ to $30R_d$.  
We further require that the dark matter density  be strictly  smaller than the visible mass density, 
 $\sigma_0 < \rho_0$.  (The data can in fact be fit equally well with the roles of dark matter and visible matter inverted.)
Here we treat $\sigma_0$, $r_s$ and $\alpha$ as free parameters.  
Together with $ \rho_0$ and $R_d$ from  the visible component,  the Dark Matter model 
utilizes five free parameters.
All in all the rotation velocity of a test star is given by 
\begin{equation}  \label{eq: eomDM}
\frac{v^2}{r} =   F_{*}  +  F_{DM} 
\end{equation}  
in the Dark Matter model.

\section{Analysis}
\label{sec: analysis}
The rotation velocity of a given test star is solved from equations  
(\ref{eq: eom_string}) and (\ref{eq: eomDM}), respectively, 
for the string and the dark matter model.  
The best fit parameters of each model are obtained by minimizing the $\chi^2$ functionals.%:
 %~\footnote{Notice that in the observations the distance measurement from the center of the galaxy is assumed to be exactly.  The uncertainty is attributed, instead,  to the velocity measurements.  During fitting, however,  we discovered that uncertainty in determine the ``center of the galaxy'' affects the quality of the fit significantly.  According to both models the rotation velocity at the ``center'' of the galaxy should be exactly zero.  If we could shift some data by a linear translation to make the zero velocity point coincide with the $r=0$ point by hand, we would have obtained much lower $\chi^2$ values for both models.  Therefore this linear shift  is better attributed to the error in distance determination.}
  We obtained our rotation curve  data of  the twenty-two galaxies in the SINGS sample from the FaNTOmM website.

In  Figure~1 rotation curve of  galaxy NGC2403 fitted using string model (left) and   NFW profile (right) are  plotted side by side for comparison. 
Squares with error bars are observational data.  Theoretical predictions are indicated by the solid lines with stars in the string fit (left) and with triangles in the NFW fit (right).  The string model clearly gives a better fit.
\begin{figure}
\label{fig: ngc2403} 
$
\begin{array}{ll}
\includegraphics[width=1.7in]{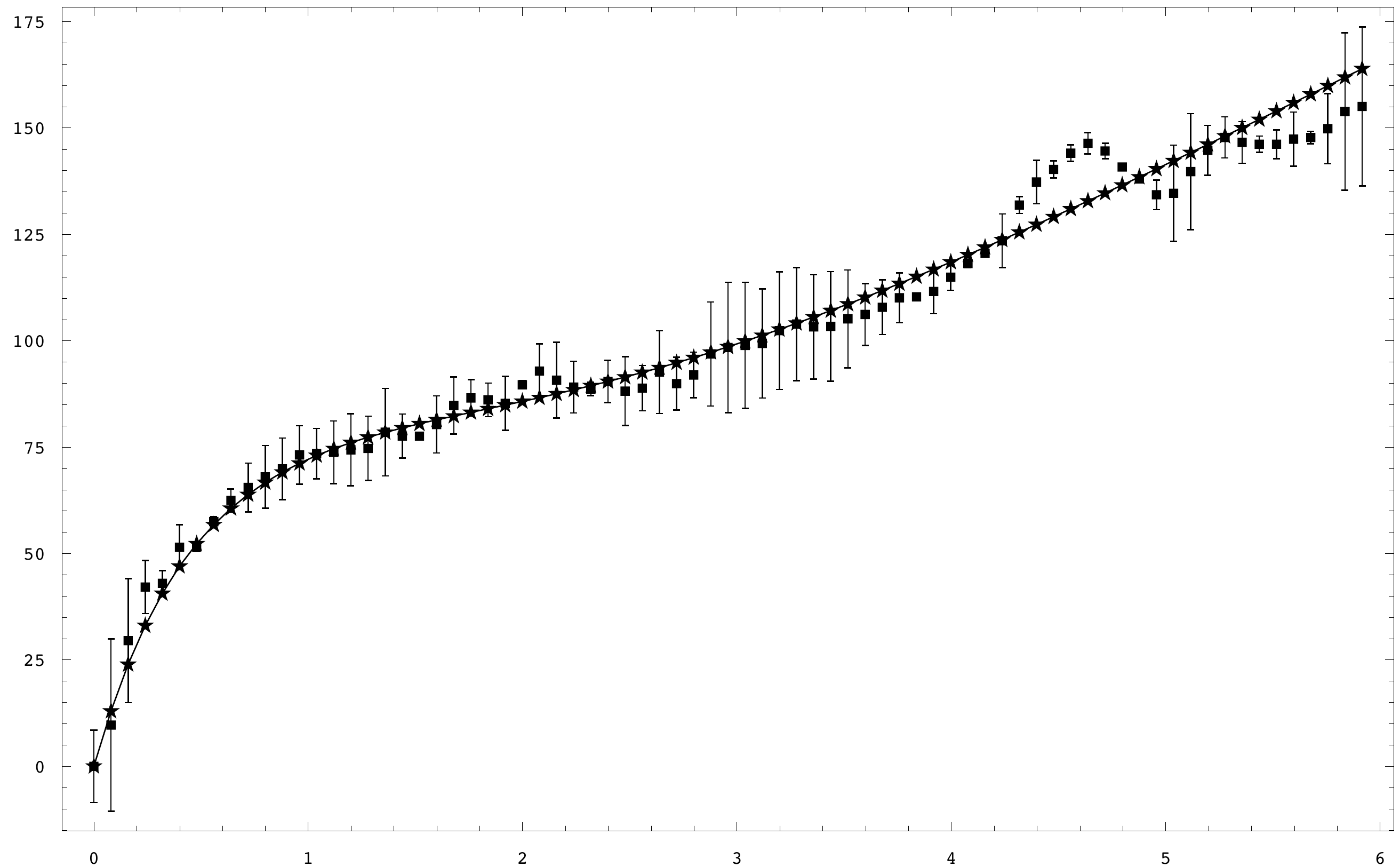} &
\includegraphics[width=1.7in]{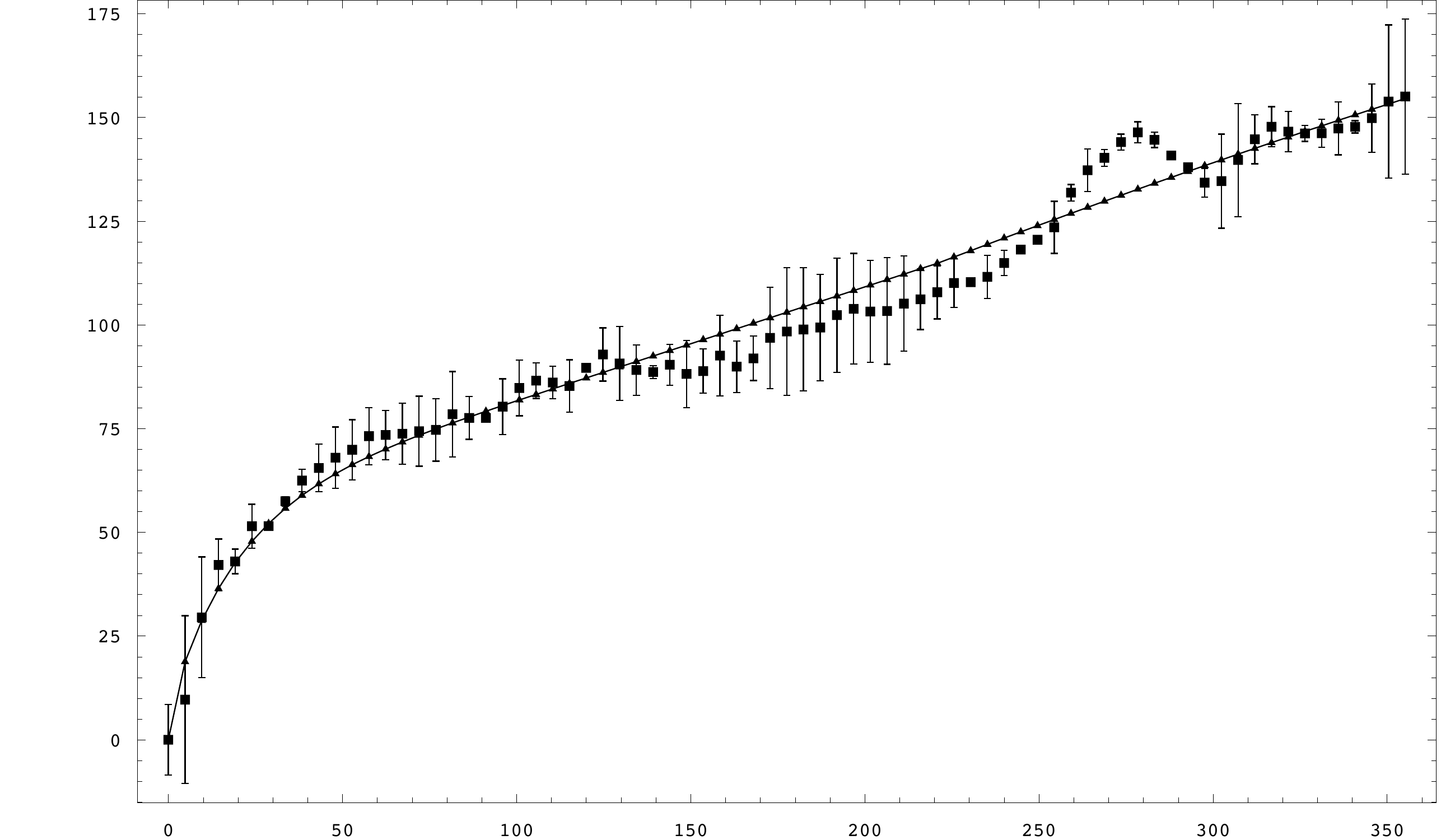} 
\end{array}
$
\caption{Rotation curve of NGC2403 fit with the  string model  (left) and with the dark matter model (right).  
The $\chi$-squared value per degree of freedom from string fit is $4.304$ while that from dark matter model is $4.515$.  
The X-axis is in $kpc$ and Y-axis in $kms^{-1}$.  
}
\end{figure}  
The $\chi$-squared value, per degree of freedom, from  the string  fit is $4.304$ whereas that  from  the NFW  fit is $4.515$.
Overall  string model gives a $\chi$-squared value  of  $1.656$ averaged over the 22 galaxies while the dark matter model gives $1.594$.  So we can see that the NFW profile  fits marginally better  at a  price of   two more  free parameters.

After we obtain the best fit values for the  free parameters we can compute the (total) masses for the galaxies.  
\paragraph*{String Model:}  For this model there is only visible matter whose mass can be straightforwardly computed by integrating (\ref{eq: vdKS}) with the best fit values of $R_{d}$ and $\rho_{0}$ for each galaxy.
\paragraph*{NFW profile:}
Matter in this model consists of the visible matter, same as that in the string model,  and the dark matter which  assumes the generalized NFW density profile~(\ref{eq: NFW}). The NFW profile gives divergent mass if the radius is integrated to infinity.  We therefore adopt the usual cutoff and compute the mass only up to the virial radius within which the average density is 200 times the critical density for closure.

\subsection*{Visible Mass to Light vs B-magnitude}
Using the measured B-band absolute magnitudes we  compute the visible mass to light ratios for the galaxies.  In string model these ratios  fall between $0.11\sim 5.6$ and centred around $1$.  The same ratios from the NFW model   span five  orders of magnitude (see Fig.~2) 
%\ref{fig: mass-to-light})
with a lot of them falling much below $1$.  
For the NFW model we also compute the percentage  of  baryonic matter in the total mass.  According to the CDM paradigm this number should be around $10\%$.  However the actual results come with a wild scatter.  The  scatter in the mass-to-light ratios and the baryon fractions clearly indicate that NFW profile is not capturing the underlying physics correctly.  
\begin{figure}[h!]  
\label{fig: mass-to-light}
$
\begin{array}{rl}
\includegraphics[width=1.6in]{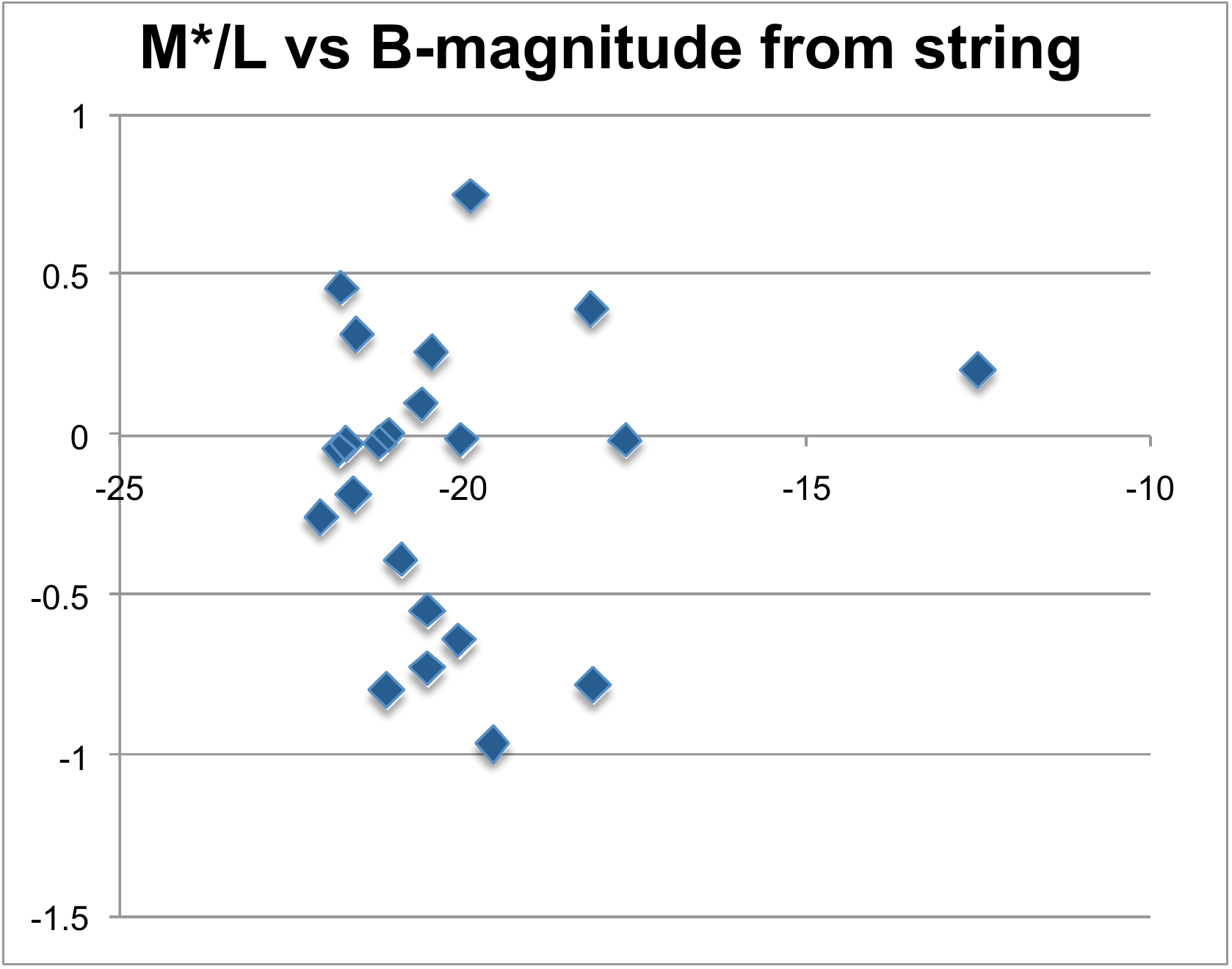} &
\includegraphics[width=1.6in]{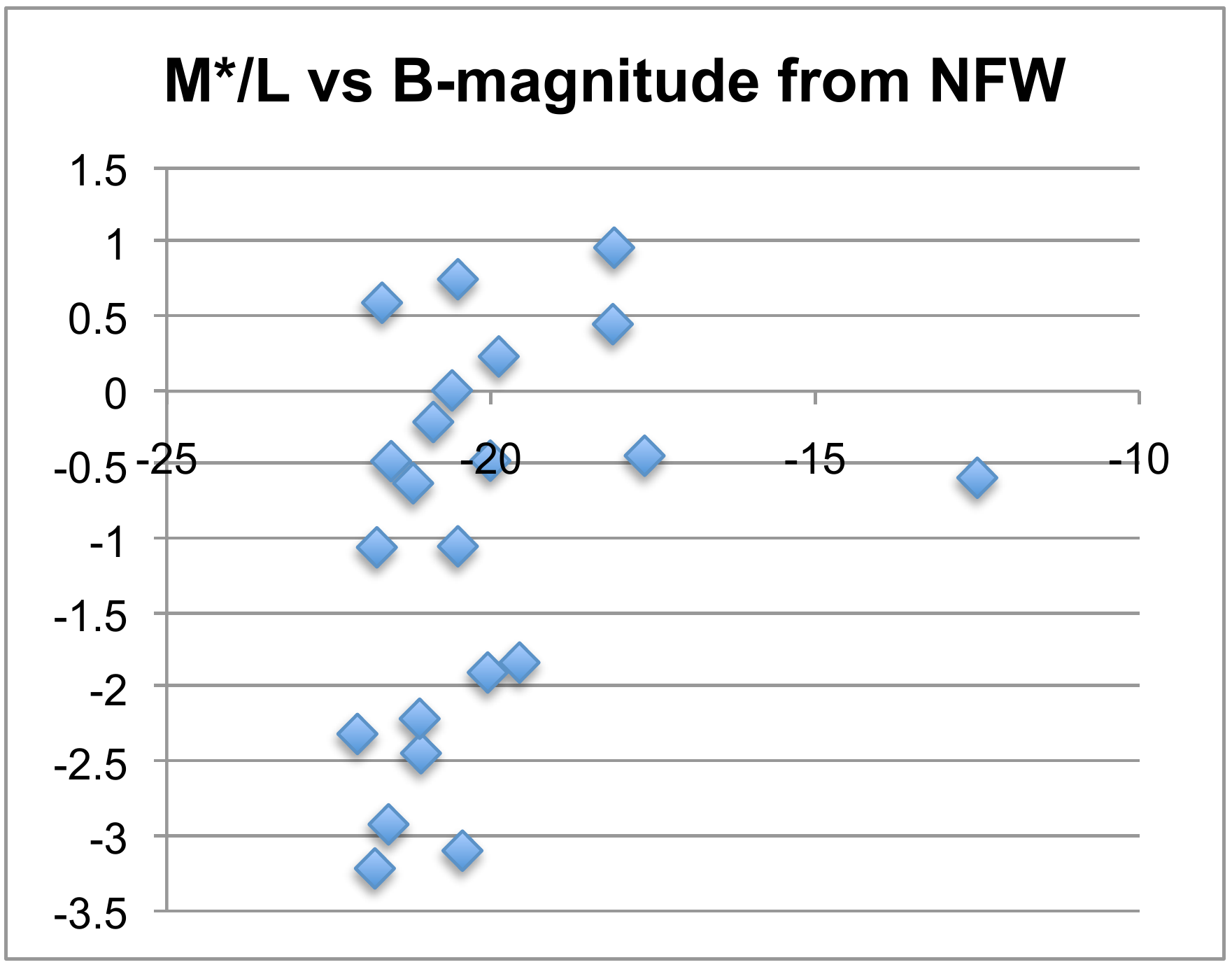} 
\end{array}
$
\vspace{-0.1in}
\caption{The visible-mass-to-light ratios derived from the best fit values and the measured B-band luminosity of the 22 galaxies in string model (left) and in dark matter model (right).  Both plots are in log scale for easy comparison.}
\vspace{-0.1in}
\end{figure}

\subsection*{Tully-Fisher relation}

A Tully-Fisher relation  can be derived from the string model which relates the rotation velocity in the ``flat'' region of the rotation curves to the product of the total luminous mass, $M_{*}$, and the parameter, $\Omega\,$,
\begin{equation}
\label{eq: TF}
v_{0}^{3} = GM_{*} \Omega~.
\end{equation}
From the equation of motion~(\ref{eq: EOM}) we  solve for $v$,
\begin{equation} \label{eq: v}
v = \Omega\, r\, + \, \sqrt{ \Omega^{2}\, r^{2}\, + F_{*} \, r }~.
\end{equation}
We then look for a balance of falling Newtonian attraction and rising Lorentz force, resulting in 
$\frac{ \partial v} {\partial r} \sim 0$.  Because we know that the turning point is at $r \sim 2.2 r_{d}$,  setting $\frac{ \partial v} {\partial r} \sim 0$ yields  a relation between  $r_{d}$ and  $\Omega$:
\begin{equation} 
\label{eq: omega}
8 \, \Omega^{2}\, \sim \frac{GM_{*}}{r_{d}^{3}}~.
\end{equation}
Inside the orbit  $r \sim 2.2R_{d}$ lies  most of  the visible mass. We have therefore used the point-mass approximation in computing $F_{*}$ and $\frac{ \partial F_{*}} {\partial r}$.  
Upon substituting~(\ref{eq: omega}) into~(\ref{eq: v}) to eliminate $r_{d}$   our Tully-Fisher relation  follows.  The string model therefore provides a dynamic origin of this well-tested rule of thumb.
\begin{figure}[h!]  
\label{fig:Tully-Fisher}
%$
%\begin{array}{cc}
%\includegraphics[width=1.6in]{luminosity-velocity.pdf} &
\includegraphics[width=2.1in]{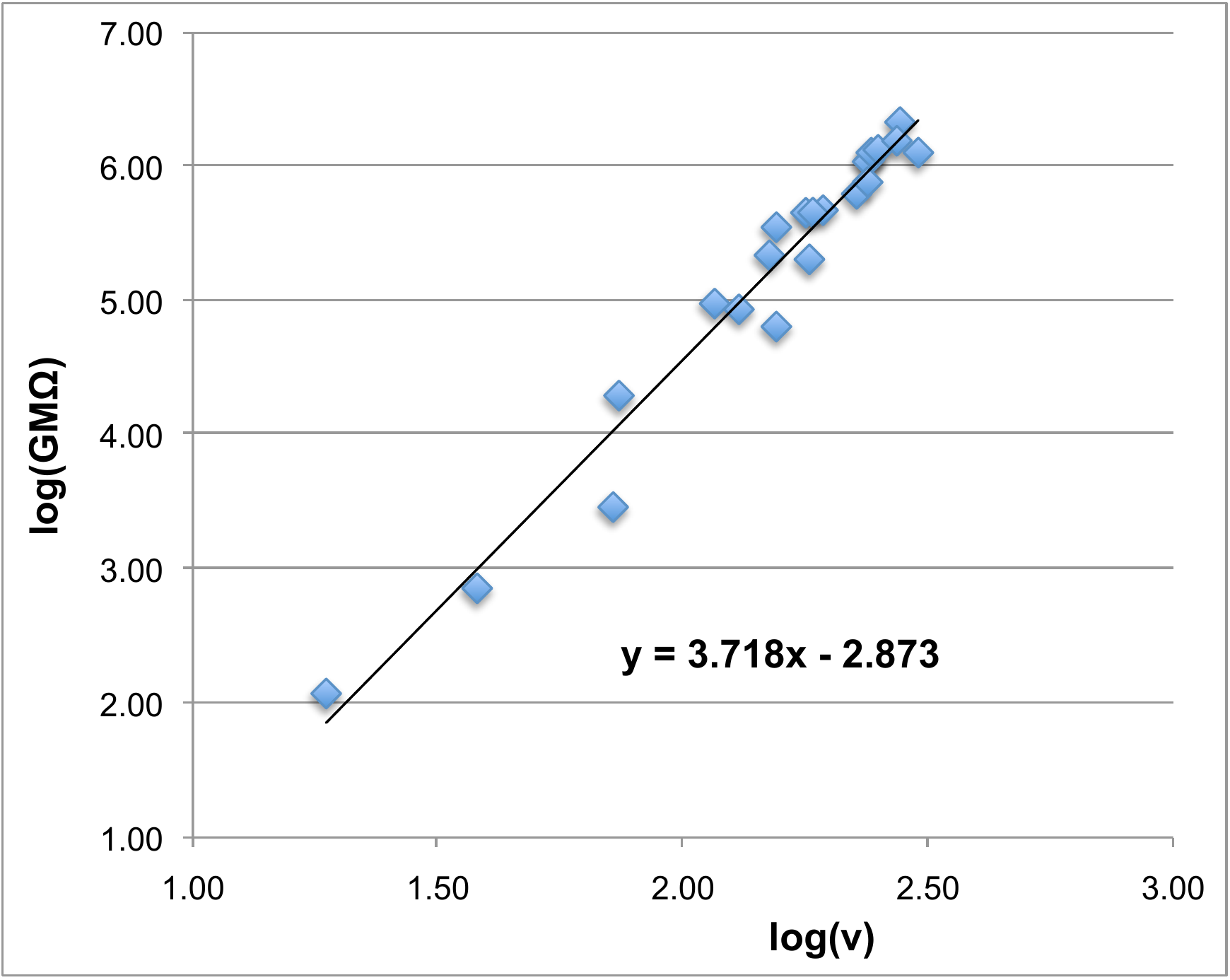}
%\end{array}
%$
\vspace{-0.1in}
\caption{The luminous mass and velocity relation of the 22 galaxies fit  by the Tully-Fisher  relation derived from the string model.}
%\vspace{-0.1in}
\end{figure}
We  plot our best fit values of $GM_{*} \Omega$ against $v$ in Fig.~3.%\ref{fig:Tullly-Fisher}.  
The representative velocity, $v_{0}$, is selected to be the {\it{maximal}} observed velocity in the  entire curve for each galaxy, to eliminate man-made bias. 
 This no doubt introduces more scatter than necessary.
Despite that  the data obey the  relation reasonably well.

\section{Discussion}

The original appeal of the NFW profile based on the ideas of hierarchical clustering was its universality.  
One simple NFW profile was expected to explain structure formation, rotation curves of galaxies, giant or dwarf, from high to low surface brightness.  
This promise has been undermined by the  cusp and core debate in dwarf galaxies  as well as  in the low surface brightness galaxies (see for example~\cite{2001MNRAS.325.1017V}).
The fact that light does not follow dark matter--well established by detailed observation and analysis in the Milky Way (see \cite{Gilmore1997} for a summary),   in addition to  a clear deficit of satellite galaxies in MW  have only served to thicken the plot.   
While we are not claiming that our string toy model can answer all these questions in one stroke we merely show that it pans out  just as well as  the Dark Matter model in  fitting the galaxy  rotation curves while using  two fewer parameters.  
Moreover by tuning the ratio of the strength of the string field  to  stellar mass density   galaxies with a wide range of surface brightness and sizes can be accommodated.   We have one dwarf and several LSB galaxies in our sample.
At the same time the model, based as it is on a tractable physical principle consistent with laws of mechanics and special relativity, does not suffer from the arbitrariness and puzzling inconsistencies of MOND. 

In order to describe  a universal galaxy rotation curve~\cite{Rubin1985, PSS} one at most needs three parameters--to 
specify the initial slope, where it bends, and the final slope.  
Any more parameter is redundant.  
In this regard the string model utilizes just the right number of them.  
The fact that it fits well on par with  the dark matter model which employs two extra parameters should be taken seriously.  
However one should guard against reading too much into the game of fitting.  For example,  
one cannot obtain a {{\it unique}} decomposition of the mass components of a galaxy using the rotation curve data alone, a difficulty  having been encountered  in the context of  comparing different dark matter halo profiles.  
Acceptable fits (defined as $\chi < \chi_{min} +1$~\cite{Navarro:diskgalaxies}) can be gotten with dark matter alone without any stellar matter in CDM model.  And the roles of dark matter and stellar matter can be completely reverted in the fitting routine.   
The physical difference is,  on the other hand, dramatic.   
This  degeneracy is less severe in the string model in the sense that  string field cannot be completely traded off in favour of stellar matter, or  vice versa. 
However there still exists  a range of values for $R_{d}, \rho_{0}$ and $ H$,  where  ``acceptable'' fits can be obtained.  
Therefore given the quality of the presently available data  rotation curve fitting alone
cannot distinguish between  dark matter and  string field in galaxies.

However precision measurements extended to   radii $r\sim 20R_{d}$ can  distinguish  string model from the other models: a  gently rising rotation curve  in this region is a signature prediction of this  string toy model.
At this moment we are, nevertheless, encouraged by this inchoate results to pursue further.  
In a separate article we shall subject our string model to other reality checks, and we shall report on how this simple string model accounts for gravitational lensing which is often quoted as another strong evidence for the existence of dark matter at intergalactic scales.

At this point it is worth mentioning that a critical  reanalysis of  available data on velocity dispersion of F-dwarfs and K-giants in the solar  neighbourhood,  with more plausible models,  performed  by Kuijken and Gilmore concluded that  the data  provided no robust evidence for the existence of any missing mass associated with the galactic disc in the neighbourhood of the Sun~\cite{KuijkenGilmoreIII}.   Instead a local volume density of  $\rho_0= 0.10M_{sun} pc^{-3}$ is favoured,  which agrees with the value obtained  by star counting.  Dark matter would have to exist outside the galactic disk  in the form of  a gigantic halo. 
Their pioneer work was later corroborated  by~\cite{FlynnFuchs, Crezeetal, Pham, HolmbergFlynn} using  other sets of A-star, F-star and  G-giant data.  
Note that this observation can be nicely explained by our model as the field only affects the centripetal motion on the galactic plane  but  does not affect the motion perpendicular to the galactic plane.  

We  presented a simple string toy model with only one free parameter and we  showed that  it can fit the galaxy rotation curves equally well as the dark matter model with the  generalized NFW profile.   The latter employs two more free parameters compared with the string model.   The string model respects all known principles of physics and can be derived from first principle using  string theory,  which in turn unifies gravity with other interactions.  Our model has an unambiguous prediction concerning the rotation dynamics of satellites and stars far away from the center of the (host) galaxy.   The ability to test the validity of string theory as a description of low energy physics makes the exercise worthwhile.   
\vspace{-0.25in}
\section*{*~~~~~~*~~~~~~*}
\vspace{-0.15in}
We wish to thank Hong-Jian He, Tan Lu, Wei-Tou Ni, Fan Wang, Hong-Shi Zong,  and in particular 
Shude Mao, Konstantin Savvidy, and Frank van den Bosch for very useful discussions.    
We thank Maurice van Putten for providing reference~\cite{KuijkenGilmoreIII}.

\bibliographystyle{apsrev}
%\begin{thebibliography}{99}
%\bibliography{cls}
%\input{csx_ref}
\bibliography{prl}

\end{document}